\documentstyle[preprint,aps]{revtex}
\input epsf.tex
\def\DESepsf(#1 width #2){\epsfxsize=#2 \epsfbox{#1}}
\begin{document}
\preprint{\vbox{\hbox{OITS-591}\hbox{UCRHEP-T154}\hbox{hep-ph/9510441}}} 
\draft
\title {Mutual consideration of
$b\rightarrow s\gamma$ and $\mu\rightarrow e\gamma$ in supersymmetric SO(10)
grand unification} 
\author{\bf T. V. Duong $^{\ast} $, B. Dutta$^{\dagger} $ and E.
Keith
$^{\ast} $ }\address{$^{\ast} $ Department of Physics, University of
California, Riverside, CA 92521\\$^{\dagger} $ Institute of Theoretical
Science, University of Oregon, Eugene, OR 97403}
\date{February, 1996}
\maketitle
\begin{abstract}
We compare the branching
ratios for $b\rightarrow s\gamma$ and $\mu\rightarrow e\gamma$ in terms of
constraining the parameter space in supersymmetric SO(10) grand unification
models where supersymmetry is broken softly near the Planck scale by
generationally symmetric operators. We observe two general cases. One with small
$\tan\beta =2$ and the other one with large $\tan\beta$ having third generation
Yukawa coupling unification at the GUT scale. \newline\indent We show that for
small
$\tan\beta$ the branching ratio constraints allow only a smaller region of
parameter space for $\mu>0$ compared to $\mu<0$ for gluino mass $\alt$ 500 GeV.
  With large $\tan\beta$, we find acceptable regions of
parameter space with $\left|\mu
\right|\alt 1$ TeV only for $\mu<0$. The dominant constraint on large $\tan\beta$ 
with
$\mu >0$ parameter space is found to be given by the $b\rightarrow s\gamma$
branching ratio, while for large $\tan\beta$ with $\mu <0$ it is found to be
given by the $\mu\rightarrow e\gamma$ branching ratio. In many of these
acceptable regions, we find that the $\mu\rightarrow e
\gamma$ branching ratio is predicted to be within one order of magnitude of its
current experimental bound. We also show that the usually neglected gluino
mediated diagrams in
$b\rightarrow s\gamma$ can not be ignored in some regions of parameter space,
especially for large $\tan\beta$ scenarios when the gluino mass is near its
lower experimental bound.
\end{abstract}
\newpage Recently it has been demonstrated \cite{[LJ],[AS],[LAS],[cia]} that
supersymmetric grand unified theories (GUTs) with soft supersymmetry (SUSY)
breaking terms which are generated near the Planck scale can predict lepton
flavor violating processes with highly significant rates. Due to the present
experimental limits, this is shown to be especially true for $\mu\rightarrow
e\gamma$. The
$\mu\rightarrow e\gamma$ decay has been studied for both SO(10) and SU(5) grand
unification. It is found that SO(10) grand unification\cite{[GFM]} predicts the
greater rates of the two as a result of its unifying all fields of a particular
generation into one multiplet \cite{[AS]}. The decay has been studied for both
small \cite{[LJ],[AS]} and large $\tan\beta$ \cite{[cia]}, and found in both
cases to sometimes rule out parameter space and sometimes predict rates only one
or two orders of magnitude beneath the current experimental limit. In this
paper, we examine regions of parameter space for which the $\mu\rightarrow
e\gamma$ branching ratio has interesting values by which it could be a signal
for supersymmetric grand unification, and test to see if those regions are
reduced by consideration of the quark flavor violating process $b\rightarrow
s\gamma$. It has also recently been shown \cite{[we]} in the case of small
$\tan\beta$ SU(5) grand unification that with the present accuracy in the
determination of the
$b\rightarrow s\gamma$ branching ratio, one can not neglect the effects of
running the soft SUSY breaking parameters above the GUT breaking scale and
should include the gluino mediated deacy amplitude when the gluino mass is not
particuarly large. This will be found to be true with SO(10) grand unification
as well, and especially true for the large $\tan\beta$ case where sometimes the
gluino mediated amplitude can be even larger than the standard model (SM)
amplitude. We will also find that applying the constraints from both decays
simultaneously, as must be done, limits available parameter space much more
severely than one of the two constraints alone would. We will be primarilly
interested in regions for which the magnitude of $\mu$, the coefficient of the
Higgs superpotentail interaction $\mu H_1 H_2$, is $\alt$1 TeV, so 
as to avoid models requiring a fine tuning of their parameters
\cite{[AC]}. 

The prediction of significant rates for lepton flavor decay processes and the
increased importance of the gluino mediated $b\rightarrow s\gamma$ decay
amplitude are the result of the fact that in GUTs the top is unified with other
third generation fields which then also feel the radiative effects of the
relatively large top Yukawa coupling. If a universal soft supersymmetry breaking
condition exists at a scale $M_X >M_G$, then the radiative corrections in the
masses of these third generation fields' soft SUSY breaking scalar masses are
manifest as $\ln{M_X/M_G}$ \cite{[Kost]} rather than suppressed as powers of
$1/M_G$, where
$M_G$ is the GUT breaking scale. Renormalization causes the soft SUSY breaking
masses for the fields unified with the top fields to become lighter than the
other two generations of those fields, and leads to a suppression of the GIM
mechanism in processes mediated by these scalar fields. The degree to which this
occurs depends, of course, on the size of the top Yukawa coupling and $M_X$ i.e.
the larger values these have the greater the suppression of the GIM mechanism is. To
understand the subtle points of these effects and how they lead to a substancial
enhancement of the $\mu \rightarrow e\gamma$ branching ratio in the case of
SO(10) over the case of SU(5) grand unification, one must examine the flavor
changing amplitudes in the interaction basis as discussed in detail in Ref.
\cite{[AS]}. We will take $M_X =2.4\cdot 10^{18}$ GeV through out this paper.
Above the GUT scale we will use the one loop renormalization group equations
(RGEs) as for example appear in Ref. \cite{[AS]}, although we use the convention
gaugino mass $M\rightarrow -M$ in the RGEs of that reference for the tri-linear
scalar soft breaking SUSY terms $A_i$ so as to be consistent with the convention
we will use for our s-particle mass matrices \cite{[SUSY]}. We will take the
SO(10) gauge coupling beta function to be $b_G =-3$ as an ad hoc choice since
our calculation is not very sensitive to its value and we do not know the
complete field content of the GUT model. We will also use $\alpha_s (M_Z )
=0.121$ and
$M_G =2\cdot 10^{16}$ GeV with the $M_G$ scale coupling $\alpha_G = 1/23.9$. We
will calculate the parameter $\mu$ at the tree level.

Below the GUT scale we will use the one loop RGEs in matrix form in the $3\times
3$ generation space for the Yukawa couplings and soft SUSY breaking parameters
as found in Ref. \cite{[SUSY],[AM]} rather than just running the eigenvalues of
these matrices as is often done. Although doing this does not provide any new
information when
$\tan\beta =2$, when $\tan\beta$ is large it allows one to know the relative
rotation of squarks to quarks and sleptons to leptons. When $\tan\beta$ is
small, one can diagonalize both the $3\times 3$ up Yukawa matrix ${\bf
\lambda_U}$ and all of the soft SUSY breaking mass matrices at the scale $M_X$
of the universal boundary condition, and they will then always remain diagonal.
In this case, for example, the mixing between a down squark soft breaking mass
matrix and the down quark mass basis is determined by the KM matrix, which
diagonalizes
${\bf \lambda_D}$ when in the basis where ${\bf \lambda_U}$ is diagonal. On the
contrary when $\tan\beta$ is large, both the top and bottom Yukawa couplings are
large and have important effects in the RGEs. Hence, if one chooses the soft
breaking mass matrices to be diagonal at the scale $M_X$ they will no longer be
diagonal at the weak scale since both ${\bf \lambda_U}$ and ${\bf \lambda_D}$
can not be diagonalized simultaneously.

The SM expression for the $b\rightarrow s\gamma$ amplitude has been
derived in Refs.
\cite{[TI],[NG]} , and the expressions for the additional MSSM amplitudes have
been derived in Refs. \cite{[AM],[NO],[RB]}. We will use the expressions given
in Ref.
\cite{[AM]} because those expressions use the squark mass eigenstate basis
derived from the full $6\times 6$ mass matrices, as will be discussed below, and
automatically incorporates mixing between ``right-handed" down squarks and
right-handed down quarks as is inevitable with either large $\tan\beta$ or
SO(10) grand unification. We will include the W-boson, charged Higgs, chargino,
and gluino mediated amplitudes, however we will neglect the neutralino mediated
amplitude since we find this to be inconsequential in all viable regions of
parameter space. The calculation is performed in the basis where
$\bf{\lambda_D}$ is diagonal at the weak scale. In terms of the mass eigenstates
$q$ and current eigenstates $q^0$ for quarks, we use $d_{L,R}=d^0_{L,R}$ and
$u_{L,R}=Vu^0_{L,R}V_R^{\rm T}$, where $V$ is the KM matrix and $V_R$ is the
analogous mixing matrix for the relative right-handed rotation between $d$ and
$u$. In the SM $V_R$ is not of any physical signifigance, however here it will
effect the chargino-quark-squark vertex. 

Neglecting the amplitude for $b_L
\rightarrow s_R \gamma$, the leading-order QCD corrected branching ratio $B$ for
$b\rightarrow s\gamma$ is given by
\begin{eqnarray} B\left(b\rightarrow
s\gamma\right)&=&\frac{\Gamma\left(b\rightarrow
s\gamma\right)}{\Gamma\left(b\rightarrow ce\bar{\nu}\right)}B\left(b\rightarrow
ce\bar{\nu}\right) \, , \end{eqnarray} where $B\left(b\rightarrow
ce\bar{\nu}\right) =0.107$ is the experimentally determined value, 
and $\Gamma \left(b\rightarrow
ce\bar{\nu}\right) ={G_F^2 m_B^5/ 192\pi^3}\left| V_{cb}\right|^2 g \left( 
m_c/m_b\right)$ with $g \left( 
m_c/m_b\right)$ being the phase space factor and $m_c/m_b = 0.316$.   The
inclusive width for the $b\rightarrow s\gamma$ is given by
\begin{eqnarray}\Gamma\left(b\rightarrow s\gamma\right)&=&\frac{m^5_b}{16\pi}
\left|c_7\left(m_b\right)\right|^2
\end{eqnarray} The QCD corrected amplitude $c_7(m_b)$ is given as
\begin{eqnarray} c_7\left( m_b\right) =\eta^{16/23} \left[ c_7\left( M_W\right)
-{8\over 3}c_8\left( M_W\right) \left[ 1-\eta^{-2/23}\right]\right]
+\sum_{i=1}^8{a_i\eta^{b_i}} \, , \end{eqnarray} with $a_i$ and $b_i$ being
given in ref. \cite{[AJ]}, $\eta =\alpha_s(M_W) /\alpha_s(m_b)$, for which we
will use $\eta =0.548$. The present experimentally accepted range for this
branching ratio is $\left(1-4.2\right)\cdot 10^{-4}$ \cite{[PS]} at the
95$\%$ C.L.. The terms $c_7(M_W)$ and
$c_8(M_W)$ are respectively $A_\gamma$, the amplitude for $b_R\rightarrow
s_L\gamma$ evaluated at the scale $M_W$ and divided by the b-mass $m_b$ and
$A_g$, the amplitude for
$b_R\rightarrow s_L g$ also given in Ref. \cite{[AM]} divided by the factor $m_b
\sqrt{\alpha_s /\alpha}$. The effective interactions for $b\rightarrow s\gamma$
and
$b\rightarrow sg$ are given by
\begin{eqnarray} L_{\rm eff}={m_b \over 2}\left( A_\gamma
\overline{s}\sigma^{\mu\nu} P_RbF_{\mu\nu}+A_g\overline{s}\sigma^{\mu\nu}
P_RbG_{\mu\nu}\right) +{\rm h.c.}\, .
\end{eqnarray} In calculating $c_7(M_W)$ and $c_8(M_W)$, we will use the
conventional approximation of taking the complete MSSM to be the correct
effective field theory all the way from the scale $M_G$ down to $M_W$, and hence
ignore threshold corrections. 

We acknowledge that since we are working with SO(10) grand unification with $M_X 
> M_G$, that there will necesarilly be a gluino mediated contribution to 
$b_L\rightarrow s_R\gamma$, however we find this contribution to not be of much
signifigance in any of the scenarios which we consider. This is dispite the fact
that the gluino mediated contribution to $b_R\rightarrow s_L\gamma$ is often
important. Three major differences exist between these two contributions to the
branching ratio. First of all, the the gluino mediated $b_L\rightarrow
s_R\gamma$ contribution does not have an interference term with an appreciable
SM amplitude or any other appreciable MSSM amplitude. Secondly, even though the
$\tilde{g}-\tilde{b_R}-s_{R}$ and the $\tilde{g}-\tilde{b_L}-s_{L}$ vertexes
have the same mixing angles at the GUT scale due to the symmetric nature of the
10-dimensional Higgs, the ``right-handed" mixing angle is smaller than the
``left-handed" mixing angle at the weak scale. To understand this one can observe
the case of small $\tan\beta$, where the $\tilde{g}-\tilde{b_R}-s_{R}$ vertex's
mixing angle is given by $\left| V_{ts} (M_G)\right| \approx 0.03$ since
right-handed quark mixing angles $ V_{ts}^R$ essentially do not run in
the MSSM:
\begin{eqnarray} 16 \pi^2 {d \ln{\left| V^R_{ts}\right|} \over dt}
= -2\lambda_t^2  {\lambda_s
\over
\lambda_b} -
2\lambda_b^2 {\lambda_c \over \lambda_t}\, , \end{eqnarray} as can be surmised
from Ref. \cite{[Ma]}, while the
$\tilde{g}-\tilde{b_L}-s_{L}$ vertex's mixing angle is given by $\left| V_{ts}
(M_W) \right|  >\left| V_{ts} (M_G)\right|$. In fact for small $\tan\beta$ if one
uses 
$1-V^*_{tb}V_{tb}\approx 0$, the ratio of the two gluino mediated amplitudes may
be estimated as follows:
\begin{eqnarray} \left| {A_{L\rightarrow R} \over A_{R\rightarrow L}}\right|   
&\approx &
\left| {V^R_{ts} \left( M_W\right)
\over V_{ts}\left( M_W\right)}\cdot { G_2\left(
\tilde{b_L} ,\tilde{b_R}\right) - G_2\left( \tilde{b_L} ,\tilde{d_R}\right) \over
G_2\left( \tilde{b_R} ,\tilde{b_L}\right) - G_2\left( ,\tilde{b_R}
\tilde{d_L}\right)}\right|\, , \end{eqnarray} where the functions $G_2 (x_1,x_2,)$
are given in Ref. \cite {[AS]} and used for the case of gluino mediated decay in
Ref. \cite{[we]}. Thirdly, even though
$\tilde{b_R}$ and
$\tilde{b_L}$ have the same mass terms at the GUT scale, this is not true at the
weak scale. With low $\tan\beta$, $\tilde{b_L}$'s mass is driven lower than that
of
$\tilde{b_R}$. Although the opposite is true for large $\tan\beta$, there the
mixing angle for $\tilde{g}-\tilde{b_R}-s_{R}$ is further suppressed, due in
part to the nature of the boundary condition at $M_G$. In none of the examples
that we look at do we find $\left| A_{L\rightarrow R} / A_{R\rightarrow L}\right|$
to be greater than about $0.65$, and further in the regions where $A_{L\rightarrow
R} / A_{R\rightarrow L}$  is greater than $0.5$ the parameter space is ruled out
by either the $\mu\rightarrow e\gamma$ or the $b\rightarrow s\gamma$ branching
ratio being too large. On the other hand, we find the contribution to the
branching ratio of $\mu\rightarrow e\gamma$ from the widths of
$\mu_L\rightarrow e_R\gamma$ and
$\mu_R\rightarrow e_L\gamma$ to be virtually the same \cite{[AS]}.

Using notation analagous to that used for $b\rightarrow s\gamma$ in Ref.
\cite{[AM]}, we now give the expressions we use to calculate the $\mu
\rightarrow e\gamma$ branching ratio.
$\mu
\rightarrow e\gamma$ has the following effective Lagrangian: \begin{eqnarray}
L_{\rm eff}={m_{\mu} \over 2}\left( A_{1\gamma} \overline{e} \sigma^{\mu \nu}
P_R\mu F_{\mu\nu}+A_{2\gamma}\overline{e} \sigma^{\mu\nu} P_L\mu
F_{\mu\nu}\right) +{\rm h.c.}\, , \end{eqnarray} where we have included
helicities.
$A_{1\gamma}$ and $A_{2\gamma}$ are given by \begin{eqnarray}
A_{1\gamma}&=&-{{\alpha \sqrt{\alpha}}\over {2\cos^2 \theta_W
\sqrt{\pi}}}\sum_{j=1}^4\sum_{k=1}^6
\frac1{M_{\tilde{l_k}}^2}\times\\\nonumber && \left\{\left(\sqrt
2G^{jk\mu}_{0lL}\right) \left(\sqrt 2
G^{*jke}_{0lL}\right) F_2\left(x_
{\tilde{\chi_{j}^0}\tilde{l_k}}\right) -\right. \\\nonumber && \left.\left(\sqrt
2G^{jk\mu}_{0lR}-H^{jk\mu}_{0lL}\right) \left(\sqrt 2
G^{*jke}_{0lL}\right)\frac{m_{\tilde{\chi^0_j}}}{m_\mu} 
F_4\left(x_{\tilde{\chi_{j}^0}\tilde{l_k}}\right) \right\}\\\nonumber
A_{2\gamma}&=&-{{\alpha \sqrt{\alpha}}\over {2\cos^2 \theta_W
\sqrt{\pi}}}\sum_{j=1}^4\sum_{k=1}^6
\frac1{M_{\tilde{l_k}}^2}\times\\\nonumber && \left\{\left(\sqrt
2G^{jk\mu}_{0lR}\right)\left(\sqrt 2
G^{*jke}_{0lR}\right) F_2\left(x_
{\tilde{\chi_{j}^0}\tilde{l_k}}\right) -\right. \\\nonumber && \left.\left(\sqrt
2G^{jk\mu}_{0lL}+H^{jk\mu}_{0lR}\right) \left(\sqrt 2
G^{*jke}_{0lR}\right)\frac{m_{\tilde{\chi^0_j}}}{m_\mu} 
F_4\left(x_{\tilde{\chi_{j}^0}\tilde{l_k}}\right) \right\} \end{eqnarray} where
the convention $x_{ab}={m^2_a}/{m^2_b}$ has been adopted and $M_{\tilde{l_k}}$ are
the slepton mass eigenstates. The functions
$F_2\left(x\right)$ and $F_4\left(x\right)$ are given in Ref. \cite{[AM]} and
\begin{eqnarray} G^{jki}_{0lL}&=&-1/2[Z_{j1}+\cot\theta_W
Z^*_{j2}]\Gamma^{ki}_{lL}\\\nonumber
G^{jki}_{0lR}&=&-Z_{j1}\Gamma^{ki}_{lR}\\\nonumber
H^{jki}_{0lL}&=&Z_{j3}\left(\Gamma_{lL}\hat{Y}_l\right)^{ki}\\\nonumber
H^{jki}_{0lR}&=&Z_{j3}\left(\Gamma_{lR}\hat{Y}_l\right)^{ki}\, ,\\\nonumber
\end{eqnarray} where $\hat{Y}_l\equiv
{\rm diag}\left(\lambda_e,\lambda_\mu,\lambda_\tau\right)/\left( g'\right)$,
$Z$ is the
$4\times 4$ neutralino mixing matrix on the $\left(\tilde{B},
\tilde{W_3},\tilde{H^0_1},\tilde{H^0_2}\right)$ basis, the fact the $4\times 4$
neutralino mass matrix can have negative eigenvalues must be taken into account
\cite{[negm]}, and $\tilde{l}_{L,R} =\Gamma^{\dagger}_{lL,R} \tilde{l}$. As
explained in Ref.
\cite{[AM]}, the terms proportional to the function $F_2$ are small compared to
the terms proportional to $F_4$ in SO(10) grand unification. For small
$\tan\beta$, the terms proportional to $F_2$ provide about a 5-percent correction
to the branching ratio. As one would expect, when
$\tan\beta$ is large we find very little difference between the branching ratios
predicted for the two possible signs of $\mu$. The Higgsino-gaugino mediated terms
are also found to provide a small correction, about 5-percent, when $\tan\beta
$ is large. The width is then given by:
\begin{eqnarray}
\Gamma\left( \mu\rightarrow e\gamma\right) ={m_\mu^5 \over16 \pi^2} \left(
\left| A_{1\gamma}\right|^2 + \left| A_{2\gamma}\right|^2\right)\, .
\end{eqnarray} The experimental upper limit on the $\mu \rightarrow e\gamma$
branching ratio is $4.9\cdot 10^{-11}$ at the 90 $\%$ C.L. \cite{[PP]}. 

The
$6\times 6$ slepton or squark mass matrix can be written in the $3\times 3 $
forms having the submatrices $M_{LL}$,$M_{LR}$ and $M^2_{RR}$ as
follows:\begin{eqnarray}
\left(\begin{array}{cc} M_{lL}M_{lL}^{\dag} & M^2_{lLR} \\ M^{2\dagger}_{lLR} &
M_{lR}^{\dag}M_{lR}\end{array}\right) \, . \end{eqnarray} At a scale $M_X^2$
near the Planck scale, the $3\times3 $ blocks acquire the form:
\begin{eqnarray} M_{lL}M_{lL}^{\dag}=m^2_0{\bf 1}\, ,\,
M_{lR}^{\dag}M_{lR}=m^2_0{\bf 1}\, ,\, M^2_{lLR}=0\, , \end{eqnarray} where we
have taken the $M_X$ scale tri-linear soft SUSY breaking scalar coupling $A^0=0$.
For simplicity, we will take
$A^0 =0$ in all of the cases that we consider. These $3 \times 3$ mass matrices
must be run down to the weak scale where one must add the mass terms that arise
from weak scale gauge symmetry breaking including the weak scale $D$ terms. The
weak scale soft SUSY breaking $3\times 3$ mass matrices may be expressed in
terms of the  universal soft breaking parameters. Hence, the weak scale
$3\times 3$ matrices may be written as:
\begin{eqnarray}\left(M_{lL}M_{lL}^{\dag}\right)_{ij}&=&a_{ij}M^2_{10}+
b_{ij}m^2_0+\left(M_{l}M_{l}^{\dag}\right)_{ij}+\\\nonumber &&
M^2_Z\cos2\beta\left({1\over 2}-\sin^2\theta_W\right)\delta_{ij}\, ,\\\nonumber
\left(M^2_{lLR}\right)_{ij}&=&\left(c_{ij}M_{10}+\mu\tan\beta
\right)\left(M_l\right)_{ij}\, ,\\\nonumber \left(
M_{lR}^{\dag}M_{lR}\right)_{ij} &=&d_{ij}M^2_{10}+e_{ij}m^2_0+
\left(M_{l}M_{l}^{\dag}\right)_{ij}+\\\nonumber
&&M^2_Z\cos2\beta\left(-1\sin^2\theta_W\right)\delta_{ij}\, ,\\\nonumber
\end{eqnarray} where $M_{10}$ is the gaugino mass at the $M_G$ scale and the
dimensionless coefficients are determined by the numerical solutions of the RGEs
\cite{[AM]}. We can write the up and the down squark sectors in the same
$6\times 6$ matrix form with the $3\times 3$ blocks looking like those shown
above.

Since we are working with the  SO(10) unification group, the above equations for the
low energy mass matrices may be modified by additional D-terms which are
generated at the GUT breaking scale $M_G$. A D-term contribution may appear when
the rank of the group is reduced due to the gauge symmetry
breaking\cite{[HK],[FH],[KMY]}. In the case of SO(10) breaking to the SM gauge
group, the rank is reduced by one and the broken generators constitute a U(1)
subgroup. In this case, the D term contribution can be described by one
additional by parameter, which we will refer to as $m_D^2$. The scalar masses
get modified
\cite{[cia]} at the unification scale as follows: \begin{eqnarray} m^2_{H_1,H_2}&=&m^2_{10}+
\left\{
\begin{array}{r} 2\\ -2
\end{array}
\right\}m^2_D \, ,
\label{eq:mH0}
\\ m^2_{Q,U,E,D,L}&=&m^2_{16}+
\left\{
\begin{array}{r} 1\\ 1\\1\\ -3\\-3
\end{array}
\right\}m^2_D \, .
\label{eq:mF0}
\end{eqnarray} These additional terms lead to an additional term proportional to $m_D^2$
in the expansion of each
$3\times 3$ squark or slepton mass matrices $M_LM^\dagger_L$ or $M_R^\dagger
M_R$ at the low scale. One must remember that when $X_Y\equiv \sum_i { Y_i
m_i^2}
\neq 0$ at the $M_G$ scale one must add an additional term to the soft breaking
scalar mass RGEs which scales in a simple fashion (see Ref. \cite{[cia]}). In
SO(10) models, $X_Y (M_G) =M_{H_2}^2(M_G)-M_{H_1}^2(M_G)$.

In large $\tan\beta$ scenerios due to the effect of the bottom Yukawa coupling now
being large as well as the top Yukawa coupling, it is very difficult to make
($m^2_{H_2}+\mu^2$) negative so as to break the electroweak symmetry and yet
keep the pseudoscalar mass-squared 
$m_A^2 =m^2_{H_1} + m^2_{H_2} +2\mu^2$ positive, however the introduction of the
additional GUT scale D-terms can solve that problem. The requirement to keep the
psuedoscalar mass-squared positive gives a lower bound on the possible size of
$m^2_D$ for a choice of all of the other parameters. An upper bound is imposed
on $m_D^2$ through the requirement that all of the down squark and slepton mass
eigenstates remain positive. For small $\tan\beta$ since only the top Yukawa
coupling is large, we do not have the above mentioned problem. In this case, the
lower bound on $m_D^2$ is 0. For simplicity and because the presence of a
nonzero $m^2_D$ raises the value of $|\mu |$, we will take $m_D^2 =0$ for
$\tan\beta =2$, but we will invoke the parameter in the full allowed range for
large $\tan\beta$.

In the high $\tan\beta$ cases, we assume the complete unification of the third
generation Yukawa couplings. This means that both MSSM Higgs doublets come from
the same $10$-dimensional representation Higgs field. This would seem to
indicate that all of the Yukawa coupling matrices are identical at the GUT
scale, which can not produce a realistic fermion mass spectrum and quark mixing
parameters at the low scale. However if one assumes that only the $(3,3)$
entries of the Yukawa matrices are direct couplings to the low mass Higgs field
and the other couplings arise from non-renormalizable operators involving super
heavy fields, then it is possible. Maximally predictive Yukawa textures have
been developed for this case which use only four operators are as follows 
\cite{[ADHRS]}: \begin{eqnarray}
\lambda^i&=&\left(\begin{array}{ccc} 0 & \acute{z_i} C_i & 0\\ z_i, C_i & y_i E
e^{i\phi} & \acute{x_i}B\\ 0 &x_iB & A
\end{array}\right) \, ,
\end{eqnarray} where $x_i,y_i,z_i$ and $\acute{x_i},\acute{y_i},\acute{z_i}$ are
Clebsch factors. To determine the four magnitudes, the phase, and $\tan\beta$,
one should use the six best determined low energy parameters as discussed in
Ref. \cite{[ADHRS]}. From this reference, we choose to use for our example the
ansatz which is its Model 6, which appears to give the most comfortable fit to the
low energy data  amongst its models. With that ansatz we will look at
two scenarios corresponding to two different values of $A\approx \lambda_t
(M_G)$. In the first example, case (i), $A=1$ and $\tan\beta =57.15$ and this
gives us the pole mass $m_t =182$ GeV and the running mass $m_b=4.43$ GeV. In the
second example, case (ii), $A=1.18$ and $\tan\beta =58.86$ and this gives us the
pole mass $m_t =184$ GeV and the running mass $m_b=4.35$ GeV.  In running the
single third generation Yukawa coupling
$A$ and soft SUSY breaking parameters between $M_X$ and $M_G$, we make
the simplifying assumption that $\lambda_t$ is the only large Yukawa coupling in
the GUT scale model, and that hence the soft breaking matrices remain diagonal
as one runs them down from $M_X$ to $M_G$. 

We also assume  for low
$\tan\beta$ that $\lambda_t$ is the only large tri-linear coupling in the GUT
scale superpotential, and at the scale $M_G$ we take ${\bf \lambda_U}$ and the
scalar matrices to be diagonal, and take ${\bf \lambda_D}={\bf \lambda_E}
={\bf V}^* {\bf \lambda_D}^{\rm diag} {\bf V}^\dagger$ with all parameters
evaluated at $M_G$. We also always take the tri-linear soft breaking parameter
matrices ${\bf A_{U,D,E}} (M_G)$ to be diagonal, and take the tri-linear
couplings to be given by the symmetric combination $\{{\bf A_{U,D,E}},{\bf
\lambda_{U,D,E}}
\} /2$ at $M_G$. (See Ref. \cite{[AS]})  In the low
$\tan\beta$ scenario as well as in both of the large
$\tan\beta$ examples, our KM matrix is typically described by the following four
parameters:
$|V_{us}|=0.22$, $\left| V_{cb}\right| =0.44$, $\left| V_{ub}/V_{cb}\right|
=0.7$, and the Jarlskog CP violation parameter
$2.8\cdot 10^{-5}$. This provides the ratio, $\left| V_{ts}^*V_{tb}/V_{cb}\right|^2
=0.96$, relevant to the
$b\rightarrow s\gamma$ branching ratio.

Now beginning with low $\tan\beta =2$, we will discuss the results. We use
$\lambda_t (M_G) =1.25$, which gives the top quark the pole mass $m_t =176$ GeV. We
forego the
$M_G$ scale constraint $m_b=m_\tau$ and give the b-quark a $\overline{\rm MS}$
running mass of
$4.35$ GeV. We display the
$\mu\rightarrow e\gamma$ rate as
$\log_{10} {\left(  B/4.9\cdot 10^{-11}
\right)}$.  We plot both branching ratios as functions of the
$M_X$ scale parameter
$m_0$ for fixed values of the $M_G$ scale gaugino mass $M_{10}$, and also as a
function of $M_{10}$ for fixed values of $m_0$, in the Figs 1 and 2. The
signifigance of $M_{10}$ for the weak scale gaugino masses is that the gaugino 
mass 
$M_i= {\alpha_i} M_{10}/{\alpha_G}$. As for the other parameter, the universal soft
SUSY breaking scalar mass  $m_0$ is close to the low scale ``right-handed"
s-electron mass, and differs from it only by the renormalization effects of GUT
scale gaugino loops and bino loops below the GUT scale.    

For negative
values of
$\mu$ (Fig.1), 
$\mu\rightarrow e\gamma$ (Figs 1b.and 1e) rules out the smaller values of the
gaugino masses although they are allowed by the
$b\rightarrow s\gamma$ branching ratio (Figs 1a, 1d). For any value of $m_0$,
$M_{10}$ needs to be around 100 GeV for the allowed regions. Also large values of
gaugino masses and $m_0$ tend to increase the amount of fine-tuning
needed in the model, and  invite larger values of
$\left| \mu \right|$. We note that, roughly  $\left| \mu \right|\simeq
500$ GeV when $M_{10}$ is $\simeq 240$ GeV. 
We also plot, $r_{A_g}\equiv \left|
A_{\tilde{g}}/A_{SM}\right|$, the absolute value of the ratio of the gluino
mediated contribution to the  
$b_R\rightarrow s_L\gamma$ amplitude to the SM amplitude, and find that in the
regions allowed by both decays  it could be as large as 10$\%$. For example
when 
$m_0$=700 GeV and
$M_{10}$=130 GeV, the ratio is about 0.09. One also observes that, in the
$\mu\rightarrow e\gamma$ plot there are two allowed regions of $m_0$ for
each value of $M_{10}$. For example when $M_{10}$=225 GeV, we find that values
of $m_0$ between 0-200 GeV as well as values greater than 600 GeV are allowed.

For positive values of
$\mu$, the lower values of the gaugino masses are disfavored by the $b\rightarrow
s\gamma$ branching ratio (Figs 2a, 2d). The branching ratio reduces to more
acceptable values at higher values of gaugino and scalar masses. For values of
$\left| \mu\right| <800$ GeV, one also finds that the gluino diagram is never
found to contribute more than 13$\%$ than that of the SM amplitude. The
$\mu\rightarrow e\gamma$ branching ratio  plots (Figs 2b. and 2e) have two
regions of allowed
$m_0$ values for a particular value of $M_{10}$. For example when  $M_{10}$=150
GeV, we find that  values of $m_0$ between 0-160 GeV and also values greater
than 450 GeV  are allowed. 

Now we look at the results for the two previously mentioned high $\tan\beta$
cases: case (i) with $A=1$, and case (ii) with $A=1.18$. For neither case do we
find parameter space with acceptable values of the $b\rightarrow s\gamma$ rate
for positive $\mu \alt 1$ TeV. In fact, roughly $\mu $ needs to be at least 1300 GeV
to find  an acceptable rate. For this reason, we only show plots for the
cases with negative values of $\mu$, where we find the predominant constraint to
come from the $\mu\rightarrow e\gamma$ rate. When $\tan\beta$ is large ,the values
obtained for the
$\mu\rightarrow e\gamma$ branching ratio and $r\equiv \left|
A_{\tilde{g}}/A_{SM}\right|$ as a function of $\left| \mu \right|$ when $\mu >0$ are
virtually identical to those obtained  when $\mu <0$. The plots we show are
parametric plots with the parameter $m_D^2$ varying over its allowed range. As
previously stated, we have chosen to use Model 6 of Ref. \cite{[ADHRS]}. If we had
chosen a different  model of the type given in that reference, the amount of
leptonic flavor violation would differ in a predictable way, as discussed in Ref.
\cite{[cia]}. In particular, the $\mu\rightarrow e\gamma$ branching ratio is found
to be proportional to $\left( \chi_L \chi_R /3\right)^2$, where $\chi_{L} \equiv
\acute{x}_e /\left( \acute{x}_d - \acute{x}_u\right)$ and $\chi_{R} \equiv
x_e /\left( \acute{x}_d - \acute{x}_u\right)$. In Model 6, $\left( \chi_L \chi_R
/3\right)^2 =0.36$, however in all nine models the factor ranges from 0.1 to 10.
Hence, in our plots of $\log_{10}\left( BR(\mu\rightarrow e\gamma) /4.9
\cdot 10^{-11}\right)$ although the value of this function being $0$ corresponds to
the experimental limit in the model we use, for all nine models the experimental
limit on our plots could correspond to values as low as $-0.56$ or as high as
$1.44$.

In Fig. 3a we plot the $b\rightarrow s\gamma$
branching ratio as a function of $\mu$ with $m_0$=1000 GeV for  case
(i) . The curves are for different values of $M_{10}$.
 In Fig 3b, one
notes that values of $M_{10}$ greater than 145 GeV are disfavored by
$\mu\rightarrow e\gamma$ rate though they are allowed by the $b\rightarrow
s\gamma$ rate. To find the interesting regions for the signal of $\mu\rightarrow
e\gamma$, we need to look for the part of the curve which projects on the same
$\mu$ space as that done by the
$b\rightarrow s\gamma$ curve having the same value of $M_{10}$ and where the 
$\mu\rightarrow e\gamma$ branching ratio is within a order of magnitude beneath
the experimental value. To illustrate the gluino contribution we
also plot
$r_{A_g}$ as a function of $\mu$ for different
values of the gaugino masses in Fig. 3c. In the regions allowed by the both the
decay processes the gluino diagram amplitude is found to be same or sometimes
larger than the SM amplitude.  In fact in the interesting regions, the
gluino diagram contribution can be as much  four times greater than that of the SM
diagram. 

For case (ii) we need
to raise $m_0$ to about 2200 GeV in order to find any appreciable regions  allowed
by the  $\mu\rightarrow e\gamma$ branching ratio. For this example, in the
intersesting region allowed by both decays we have that $\mu$ lies between about
-136 and -250 GeV as shown in figs 4a and 4b. The gluino diagram contribution (Fig.
4c) is as high as only 8 percent of the SM contribution in this region. The
requirement of large gaugino mass for the existence of this parameter space is the
cause for a relatively small gluino contribution.

We now summarize the constraints on the parameter space given by considering both
decays silmutaneously.  
For small $\tan\beta$ with $\mu<0$  
one finds that, a large range of gaugino  masses for any value of scalar
mass is allowed by $b\rightarrow s\gamma$ branching ratio,  but  the 
$\mu\rightarrow e\gamma$ rate is found in general to be  more restrictive on
the lighter gaugino masses.   
In the case of $\mu >0$ with small $\tan\beta$,  
it is the $b\rightarrow s\gamma$ branching ratio which provides the
more stringent constraint and in general forbids $M_{10}$ to be less than about $180$
GeV.  
For large
$\tan\beta$ with
$\mu<$0,  
$b\rightarrow s\gamma$  allows a
larger range of gaugino masses than  $\mu\rightarrow e\gamma$ does, which tends to
rule out larger values of $\left| \mu \right|$.   
On the other hand for large $\tan\beta$ with $\mu>$0,  
$b\rightarrow s\gamma$ rules out all
the  parameter space that has $\mu\alt 1$ TeV.

In this letter, we have given first complete calculation of $b\rightarrow s\gamma$
branching ratio in SO(10) grand unification with flavor uniform soft SUSY breaking
terms introduced at the reduced Planck scale, although  with this
boundary condition the branching ratio  for  $b\rightarrow s\gamma$ has been
calculated  for SU(5)  grand unification with low $\tan\beta$ in the
literature\cite{[we]}. We have performed the calcualtion for both  large
$\tan\beta$ with
$\lambda_t =\lambda_b$ at the grand unification scale  and for  low
$\tan\beta =2$. For the purpose of comparison with the constraints provided by
muon flavor violation, we have also plotted the $\mu\rightarrow e\gamma$ rate
over exactly the same parameter space. As discussed previously in our third and
fourth paragraphs and unlike previous SO(10) calculations from the reduced Planck
scale, our  calculations use the complete
$6\times 6$ squark and slepton mass matrices and include all flavor mixing effects 
induced through the one-loop RGEs by the  MSSM yukawa coupings, which are important
for the large
$\tan\beta$ calculation since $\lambda_t =\lambda_b =\lambda_\tau$ at the grand
unification scale.

In conclusion, if one is to calculate the decay rate for
the flavor changing processes in a SUSY GUT with SUSY breaking communicated by
gravity above the GUT breaking scale in the form of soft breaking mass terms, 
it
is essential to include the GUT scale renormalization group effects. The mutual
consideration of experimental limits on hadronic and leptonic flavor violating
decays can give  strong constraints on parameter space, and gives some preference 
to
negative values of $\mu$.  One then finds there are some regions of parameter 
space
which are allowed by both the 
$b\rightarrow s\gamma$ and the $\mu\rightarrow e\gamma$ decay rates, and where   
it may soon be possible to search for the signal for new physics in the
form of $\mu\rightarrow e\gamma$.

We thank N.G. Deshpande and E. Ma. This work was
supported by Department of Energy grants DE-FG06-854ER 40224 and DE-FG02-94ER
40837.

\newpage
\leftline{{\Large\bf Figure captions}}
\begin{itemize}

\item[Fig. 1~:] {{Plots for $\tan\beta =2$ and $\mu <0$.\\
a) $b\rightarrow s\gamma$ branching ratio as a
function of $M_{10}$.\\ b)  Log$_{10}{Br\left(\mu\rightarrow
e\gamma\right)
/ 4.9\cdot 10^{-11}}$ as a function of $M_{10}$.\\ c) 
$r_{A_g}\equiv 
A_{\tilde{g}}/A_{SM}$ as a function of $M_{10}$.\\ The five curves in each of a),
b),  and c) correspond to $m_0$=0, 100, 200, 300, and 400 GeV, and are labeled with
their values of $m_0$.
\\ d)
$b\rightarrow s\gamma$ branching ratio as a function of 
$m_0$.\\ e) Log$_{10}{Br\left(\mu\rightarrow e\gamma\right)
/ 4.9\cdot 10^{-11}}$ as a function of $m_0$.\\ The three curves in both
of d) and e)  correspond to  $M_{10}$=100, 150, and 225 GeV, and are
labeled with their values of $M_{10}$}\label{fig1}}. 

\item[Fig. 2~:] {{Same as Fig. 1 caption except $\mu >0$}\label{fig2}}. 

\item[Fig. 3~:] {{Plots for the large $\tan\beta$ case (i) with A=1 and $\mu <0$.
For all plots, $m_0=1000$ GeV.\\ a) $b\rightarrow
s\gamma$ branching ratio as a function of $\mu$.\\ b) 
Log$_{10}{Br\left(\mu\rightarrow e\gamma\right)
/ 4.9\cdot 10^{-11}}$ as a function of $\mu$.\\ c)  $r_{A_g}$  as a function of
$\mu$.\\ The curves corresponds to the gaugino masses $M_{10}$=65, 105, 145, 185,
225, and 265 GeV, and are labeled their values of $M_{10}$/GeV}\label{fig3}}. 

\item[Fig. 4~:] {{4a. Plots for the large $\tan\beta$ case (ii) with A=1.18 and
$\mu <0$. For all plots, $m_0=2200$ GeV.\\ a) $b\rightarrow
s\gamma$ branching ratio as a function of $\mu$.\\ b) 
Log$_{10}{Br\left(\mu\rightarrow e\gamma\right)
/ 4.9\cdot 10^{-11}}$ as a function of $\mu$.\\ c)  $r_{A_g}$  as a function of
$\mu$.\\ The curves corresponds to the gaugino masses $M_{10}$=250, 275, 300, and
325 GeV, and are labeled by their values of $M_{10}$/GeV }\label{fig4}}.

\end{itemize}
\begin{figure}[htb]
\centerline{ \DESepsf(dd1.epsf width 15 cm) }
\smallskip
\nonumber
\end{figure}

\begin{figure}[htb]
\centerline{ \DESepsf(dd2.epsf width 15 cm) }
\smallskip
\nonumber
\end{figure}

\begin{figure}[htb]
\centerline{ \DESepsf(dd3.epsf width 15 cm) }
\smallskip
\nonumber
\end{figure}

\begin{figure}[htb]
\centerline{ \DESepsf(dd4.epsf width 15 cm) }
\smallskip
\nonumber
\end{figure}

\begin{figure}[htb]
\centerline{ \DESepsf(dd5.epsf width 15 cm) }
\smallskip
\nonumber
\end{figure}

\begin{figure}[htb]
\centerline{ \DESepsf(dd6.epsf width 15 cm) }
\smallskip
\nonumber
\end{figure}

\end{document}